\documentclass[a4paper,fleqn]{cas-dc}
\usepackage[square,numbers,sort&compress,]{natbib}

\usepackage{import}
\usepackage{tabularx}
\usepackage{xurl}

\ExplSyntaxOn
\cs_gset:Npn \__first_footerline:
  { \group_begin: \small \sffamily \__short_authors: \group_end: }
\ExplSyntaxOff 

\begin{document}
\let\WriteBookmarks\relax
\def\floatpagepagefraction{1}
\def\textpagefraction{.001}

% Short title
\shorttitle{Towards an Effective Data-Driven Decision Making in UK SMEs}    

% Short authorhttps://www.overleaf.com/project/6217701e61e3e3c3f5c26907
\shortauthors{Abdel-Rahman Tawil, Muhidin Mohamed, Xavier  Schmoor, Konstantinos Vlachos, Diana Haidar}  

% Main title of the paper
\title [mode = title]{Trends and Challenges Towards an Effective Data-Driven Decision Making in UK SMEs:  Case Studies and Lessons Learnt from the Analysis of 85 SMEs}  

% \title [mode = title]{Trends and Challenges Towards an Effective Data-Driven Decision Making:  Case Studies and Lessons Learnt from the Analysis of 85 UK SMEs}  

% Title footnote mark
% eg: \tnotemark[1]
%\tnotemark[<tnote number>] 

% Title footnote 1.
% eg: \tnotetext[1]{Title footnote text}
%\tnotetext[<tnote number>]{<tnote text>} 

% First author
%
% Options: Use if required
% eg: \author[1,3]{Author Name}[type=editor,
%       style=chinese,
%       auid=000,
%       bioid=1,
%       prefix=Sir,
%       orcid=0000-0000-0000-0000,
%       facebook=<facebook id>,
%       twitter=<twitter id>,
%       linkedin=<linkedin id>,
%       gplus=<gplus id>]
\credit{
This work was funded by the European Union under the European Regional Development Fund (The Big Data Corridor project - Project no. 12R16P00220) and match-funded by six Project Partners - Birmingham City Council, Aston University, Birmingham City University, EnableID, Innovation Birmingham and West Midlands Combined Authority. The funding source had no role in the design, execution, interpretation, or writing of the work.}

\author[bcu]{Abdel-Rahman H. Tawil}[orcid=0000-0001-9610-2412]

% Corresponding author indication
\cormark[1]

% Footnote of the first author
%\fnmark[]

% Email id of the first author
\ead{abdel-rahman.tawil@bcu.ac.uk}

% URL of the first author
%\ead[url]{}

% Credit authorship
% eg: \credit{Conceptualization of this study, Methodology, Software}
%\credit{}

% Address/affiliation
\affiliation[bcu]{organization={School of Computing and Digital Technology, Birmingham City University, Birmingham, B4 7XG},
            %addressline={}, 
            %city={Birmingham},
          %citysep={}, % Uncomment if no comma needed between city and postcode
            %postcode={B4 7XG}, 
            %state={},
            country={United Kingdom}}

\author[aston]{Muhidin Mohamed}[orcid=0000-0002-8449-5818]

% Footnote of the second author
%\fnmark[]

% Email id of the second author
\ead{m.mohamed10@aston.ac.uk}

% URL of the second author
%\ead[url]{}

% Credit authorship
%\credit{}

% Address/affiliation
\affiliation[aston]{organization={Department of Operations and Information Management, ABS, Aston University, Birmingham, B4 7ET},
            %addressline={}, 
            %city={Birmingham},
         %citysep={}, % Uncomment if no comma needed between city and postcode
            %postcode={B4 7ET}, 
            %state={},
            country={United Kingdom}}

\author[bcu]{Xavier Schmoor}[]
% Email id of the second author
\ead{Xavier.Schmoor@bcu.ac.uk}

\author[bcu]{Konstantinos Vlachos}[]
% Email id of the second author
\ead{Konstantinos.Vlachos@bcu.ac.uk}

\author[bcu]{Diana Haidar}[orcid=0000-0002-0321-3357]
% Email id of the second author
\ead{Diana.Haidar@bcu.ac.uk}

% Corresponding author text
\cortext[1]{Corresponding author}

% Footnote text
%\fntext[]{}

% For a title note without a number/mark
%\nonumnote{}

% Here goes the abstract
\begin{abstract}
The adoption of data science brings vast benefits to Small and Medium-sized Enterprises (SMEs) including business productivity, economic growth, innovation and jobs creation. Data Science  can support SMEs to optimise production processes, anticipate customers’ needs, predict machinery failures and deliver efficient smart services. Businesses can also harness the power of  Artificial Intelligence (AI) and Big Data and the smart use of digital technologies to enhance productivity and performance, paving the way for innovation. However, integrating data science decisions into an SME requires both skills and IT investments. In most cases, such expenses are beyond the means of SMEs due to limited resources and restricted access to financing. This paper presents trends and challenges towards an effective data-driven decision making for organisations based on a case study of 85 SMEs, mostly from the West Midlands region of England. The work is supported as part of a 3-years ERDF (European Regional Development Funded project) in the areas of big data management, analytics and business intelligence. We present two case studies that demonstrates the potential of Digitisation, AI and Machine Learning and use these as examples to unveil challenges and showcase the wealth of current available opportunities for SMEs.

\end{abstract}

% Use if graphical abstract is present
%\begin{graphicalabstract}
%\includegraphics{}
%\end{graphicalabstract}

% Research highlights
%\begin{highlights}
%\item 
%\item 
%\item 
%\end{highlights}

% Keywords
% Each keyword is seperated by \sep

\maketitle

% Main text
%\section{}\label{}
\section{Introduction}\label{section1}

There were 5,59 million private businesses in the UK in 2021, of which 99.9\% were Small and Medium-sized Enterprises (SMEs) with fewer than 250 employees and average sales of \pounds 758,239 and less than 36 million pounds. This includes sole traders, micro, small and medium-sized businesses. SME’s accounts for 51.9\% of the private sector turnover and 53.6\% of all private sector jobs in the UK (a total of 22.9 million jobs)~\cite{MerchantSavvy}.  This data shows the importance of SMEs for the UK economy and how powerful any steps taken to assist their rapid growth would be in boosting the economy of the country. Between 2020 and 2021, as a consequence of COVID-19 pandemic and lockdown measurements, the number of businesses in the UK decreased by 6.5\%. SMEs numbers fell across all regions and countries in the UK - the greatest fall occurs in Northern Ireland, where businesses fell by 16.6\%, London saw an 8.0\% drop, and Scotland by 7.4\%~\cite{BusinessStatistics2021}.

SMEs, and their investors are recognizing the value data provides for their business~\cite{mohamed2020trends}. Contemporary companies worldwide~\cite{chege2020influence,alvarez2016have,lee2021analysis} and typically in the UK seek data-driven based innovations not only to modernise business operations and increase their competitiveness advantage – but also to carve out new markets, meet varying government policies and numerous regulators, as well as make their businesses more sustainable~\cite{wang2020big}. About 90\% of the massive amount of data that we have today has been created in the past two years alone~\cite{laney01controlling3v}, making it the "new oil" of this digital era ~\cite{Gandomi2015BeyondTH}. Data is like crude oil, without analysis is of little use to businesses if they do not know how to process or use it. Technologically efficient companies are among those that achieve high growth rates, according to research~\cite{marcinkowski2020data, khayer2020cloud}. Data-drivenness is about building tools, abilities and more crucially a culture that acts on data. A leading factor that shapes this transformation is the data collected in databases and other repositories maintained by the business. Companies that see data as a strategic asset will thrive as data becomes a key part of their competitive advantages in the coming years. Obviously, not just any data will work; it has to be the right data (e.g., timely, accurate, clean, unbiased and most importantly trustworthy). Good data has the power to transform businesses with actionable insights required to become more productive.  Evidently, there can be subtle hidden biases in the data that can sway taking conclusions. However, cleaning and massaging data can be tough, time-consuming and expensive operations. Data scientists use data analysis techniques to develop new business models that are used to deliver, create and capture value for business growth, success and profitability. Their skills are now essential to industry transformation. The analytic value chain in a data-driven organization stimulates deeper analysis. Decision makers usually incorporate these into their decision-making processes so they can influence the direction taken by the company, and therefore add value and impact. This process transforms data into knowledge and value, which creates new income streams. Yet despite the benefits and opportunities digital technologies bring, and despite the significant uptake in recent years, many SMEs are still lagging in the adoption of digital technology, and for smaller SMEs, with 10-49 employees, the digital adoption gap has widened significantly compared to larger firms.~\cite{oecd2021digital}. 

Data analytics and digital transformation offer such businesses new opportunities, such as marketing optimization, the forecasting of demand for their products and services, and staying one step ahead in retaining and acquiring customers. A survey of 500 UK companies found that using data to improve productivity and performance was positively correlated with performance: companies with high productivity are 13\% more productive than those with low productivity~\cite{lepenioti2020prescriptive}. Many government institutions, including the EU, recognize the importance of empowering SMEs to benefit from the digital revolution and generate measurable economic benefits. This is evident from the proportion of EU funding allocated to data-related projects, big data, and data science~\cite{, ghasemaghaei2019understanding}. To increase the number of highly skilled workers in AI and data science, the UK government, the Office for Students, universities, and industry partners have established a fund of up to \pounds 24 million~\cite{GovUK2020}.

In emerging economies, Small and Medium-sized Enterprises (SMEs) are estimated to generate 60\% of employment and 40\% of Gross Domestic Product (GDP). Moreover, the proportion of SMEs that employ 66\% of the workforce in the European Union is higher~\cite{lam2017leveraging}. Despite this, SMEs in the UK are adopting big data analytics at a rate of less than 1\%~\cite{willetts2020barriers,coleman2016can}. Yet businesses in the UK are becoming more aware of the value of data-driven decision-making, and data analytics is increasing in popularity, according to recent reports. The data science industry is also poised to expand over the next few years.

This paper aims to explore trends and challenges towards an effective data-driven decision making for UK businesses and how SMEs advance their Business Models (BMs) around data to handle data-driven products and how this contributes to their innovation and performance. We present an analysis of the challenges and opportunities of digitilization, and adoption of data science and Artificial Intelligence (AI) within the UK business sector. Our analysis of 85 UK SMEs is based on case studies of SMEs located primarily in England's West Midlands who are supported in the areas of data management, machine learning, data analytics, and other related digital technologies. Our study also briefly examines how small businesses can take advantage of data-driven innovation and decision-making, while highlighting challenging areas where support for digital technology adoption is most needed. The  study  explores   digital  technology  trends,  challenges  limiting  SMEs  from  effective utilization  of  enabling  technologies  and   the  state  of  their  adoption  in  AI  technologies.  Business opportunities  and  advantages  of  adopting data analytic  technologies  and AI are  demonstrated  through several selected practical case studies in Section ~\ref{section5}.

The  multi-perspective  analysis  and  case  studies  in  this  paper  inform  the  SME  business  industry  and business  innovation and growth bodies  of  potential  challenges  and  key opportunities  in  AI usage.  As well as encouraging small businesses to derive meaningful insights from closed data (those belonging to the business) and open data (those available publicly) by taking advantage of emerging trends in machine learning, and data and analytics in the past decade.  The  research  also  highlights  areas  where  future support and funding are most needed to enable SMEs enterprise sector to embark on the digital revolution and AI adoption thereby contributing to the growth and development of this key sector in the United Kingdom.

Due to non-disclosure agreements and the need to preserve business confidentiality, we do not detail individual SMEs' support or provide any associated data except in the two case studies that were used as examples with the permission of these businesses. The study focuses on the lessons learnt from the analysis of business data and the adoption of data-driven solutions by SMEs. 

The rest of the paper is organized as follows: Section \ref{section3} recaps concepts of open data and data analytics, then in Section \ref{section4}, we briefly analyse SMEs’ data and digital  technology  trends.  Section \ref{section5}  presents  related  discussion  and  the  key  lessons  learned  while supporting and collaborating with SMEs. We provide a summary and conclusion in Section \ref{section6}.

%In this paper, we summarize the lessons learned during a 3-years ERDF project aimed at assisting SMEs in the areas of data management, analytics, Artificial Intelligence, and related digital issues. We also briefly analyse the  challenges  facing  SMEs  who  want  to  make  use  of  data-driven  innovation  and  decision-making, highlighting areas where small businesses are most in need of digital technology support. Due to non-disclosure agreements and SMEs’ business confidentiality, we do not include details of support given to individual SMEs nor any associated data. The rest of the paper is organized as follows: Section 2? recaps concepts of big data and data analytics, then in Section 3, we briefly analyse SMEs’ data and digital  technology  trends.  Section  4?  presents  related  discussion  and  the  key  lessons  learned  while supporting and collaborating with SMEs. We provide a summary and conclusion in Section 5?.

%\section{Related Work}\label{section2}
%\import{sections/}{related-work}

\section{Opening Data for SMEs}\label{section3}

\begin{figure*}[ht!]
\centering
\includegraphics[width=\textwidth]{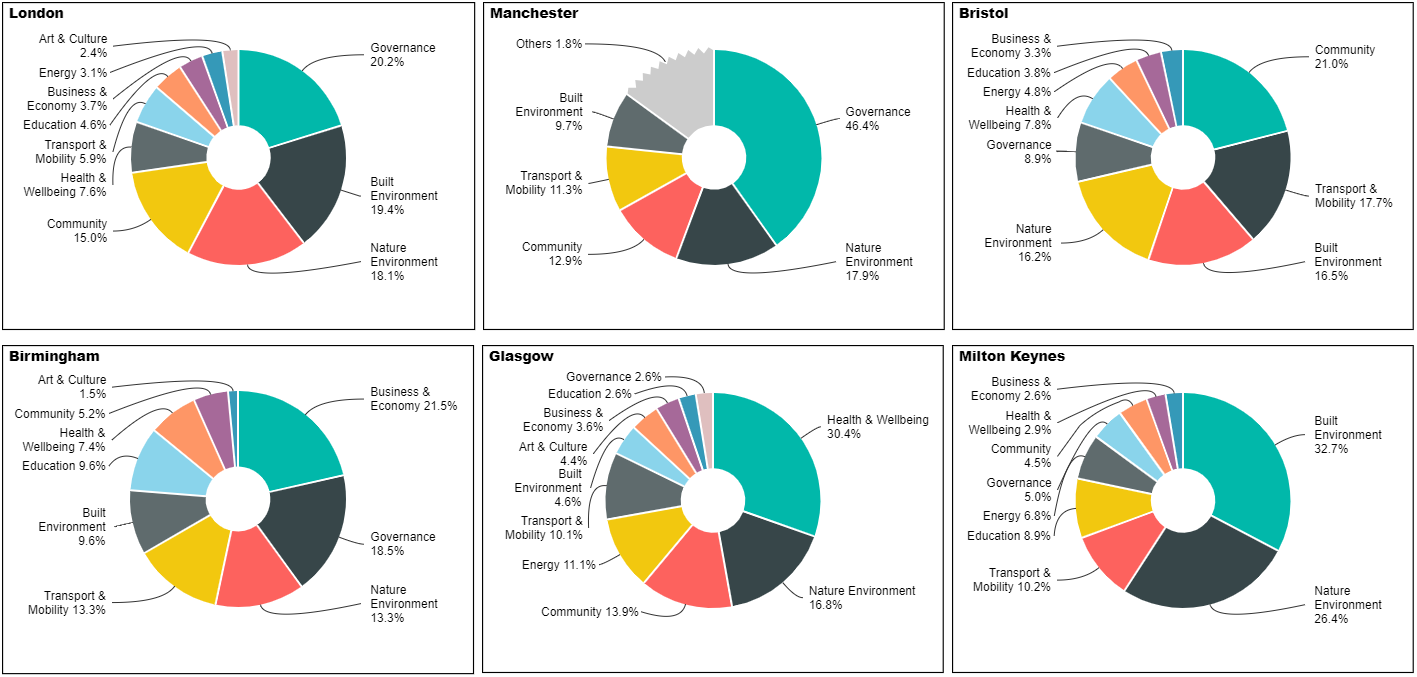}
 \caption{Open Static Datasets Distribution Across Major Themes by Six UK Cities}
    \label{fig:opendata_six_uk_cities}
\end{figure*}

In~\cite{ghahremanlou2019survey}, Ghahremanlou et al. identified several static data resources that have been established by, or are associated with six UK cities, and classified them according to these themes (domains), Figure ~\ref{fig:opendata_six_uk_cities}. 

A focus of the study is to better understand how open and shared data might impact the economy. Consider outlining how open data can be used by small and medium businesses and startups, along with how it can contribute to economic growth, new jobs, innovation, and other improvements in social and economic development aligned to smart cities. By using empirical studies to ground their analysis, we were able to identify the most important challenges facing SMEs when considering open data strategies. As well as principles that have proven, real-world applicability.

SME's can make use of collected and collated open data relatively inexpensively. Open data can be used to make innovative business decisions, or it can gain access to new data sources previously inaccessible. A data product or service can be created by combining open data with closed proprietary sources, either internally or with partners. As a result of its potential to fuel digital innovation, governments have encouraged the utilisation of open data. Researchers and policymakers alike highlight the importance of open data for SMEs in job creation and economic growth~\cite{wolf2006small}. 

The BDC project (\emph{Project name removed for double blind reviewing}) provided  us an  opportunity  to  work  with  SMEs,  supported  by Birmingham  City  Council  as  lead  project partner to motivate  and  encourage  the  use of open data. Our work contributes to data driven innovation research by analysing the main capabilities needed to overcome existing barriers to successfully take advantage of open data for SMEs support. 

We analyse the data collected from 85 business assists with SMEs participating in the BDC project and  adopting an integrated open  data-based strategies. We identify a number of core factors  that  shape open  data acquisition,  assimilation,  transformation  and  exploitation  by SMEs. Results show that without the specific capabilities delivered by our expert team, it will be difficult for these SMEs to successfully use open data, which may explain why the uptake of open data by SMEs more broadly has so far been constrained. 

As open data sources expand, SMEs and startups often find it difficult to identify and assess the usefulness of open data. It could be a result of limited time or multitasking, inability to identify available relevant data sources, a lack of understanding of open data use, or lack of understanding of how to process data. Due to time pressures, competing priorities, and limited knowledge of open data, it may be difficult to integrate this data into existing products or services. Skills are needed to develop distinctive open data capabilities by SMEs if this ‘raw material’ for the digital economy is to be fully exploited. A key message of the H2020 program guidelines on FAIR Data~\cite{landi2020fair} is that data should be "as open as possible and as closed as necessary", keeping "open" data to encourage re-use and to accelerate research, yet keeping it "closed" to preserve subject privacy. Metadata that is Find-able, Accessible, Interoperable, and Reusable (FAIR) should be developed to protect privacy rights and to meet ethical and regulatory requirements.

The Data Protection Act2018 seeks to ensure the safe and secure processing of data in order to protect individuals’ information. Personal data needs to be protected due to the fact that it can be sensitive in nature and could be used in a discriminatory way. The BDC team conducted several workshops and seminars to ensure that all members, employees and agents acting for and on behalf of Birmingham City Council (lead partner)  are aware of their obligations and responsibilities with regard to the collection and processing of personal data under the provisions of the Data Protection legislation and that it is a requirement for the BDC team to comply with all aspects and requirements of the legislation. The team also worked on several projects to demonstrate the power and benefits gain from the integration of companies proprietary data with open data.

\section{Analysis of SMEs’ Digitization Trends, Challenges and Learned Lessons}\label{section4}
%{\color{red} This section will provide an analysis of the digitalisation + challenges.... This discussion is based on the analysis of data collected from 85} 

\subsection{SMEs' Digitization Trends from Different Perspectives}
\label{SMEsDigitizationTrends}
\begin{figure}[t!]
\centering
\includegraphics[width=\columnwidth]{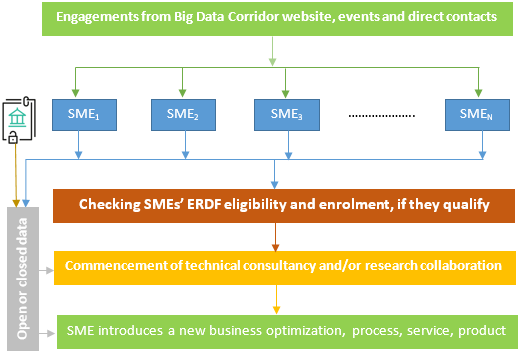}
 \caption{Lifecycle of BDC project support for SMEs}
    \label{fig:BDC-project-support-lifecycle}
\end{figure}

85 SMEs received support during the period from June 2017 to August 2019 from two academic partners of the BDC project and are used as the basis for our analysis and discussions. Figure~\ref{fig:BDC-project-support-lifecycle} depicts the BDC project support life cycle for SMEs including SME engagement, checking SME eligibility, and commencement of support which could lead to the introduction of new products or services. These SMEs were mostly located in the West Midlands region of the UK and more precisely in the Greater Birmingham and Solihull Local Enterprise Partnership (GBSLEP) area, which was the focus of the BDC project.

The SMEs of the sample had a minimum of one (the owner) and a maximum of 146 employees. The minimum turnover of the SMEs was £10,000 and the maximum reported turnover was £10 million. Companies were categorised based on their activities to 20 different sectors. Table~\ref{tab:Top-10-business-sectors} shows the top 10 business sectors based on the count of SMEs, together with the ranges for the number of employees and turnover for the SMEs of each sector.
% Please add the following required packages to your document preamble:
% \usepackage{graphicx}
%\begin{table}[<options>]
%\caption{}\label{tbl1}
%\begin{tabular*}{\tblwidth}{@{}LL@{}}
%\toprule
%  &  \\ % Table header row
%\midrule
% & \\
% & \\
% & \\
% & \\
%\bottomrule
%\end{tabular*}
%\end{table}

\begin{table*}[b!]
\centering
\caption{Top 10 business sectors among the 85 SMEs supported by BDC project.}
\label{tab:Top-10-business-sectors}
\resizebox{\textwidth}{!}{
\begin{tabular}{l l l l l }
\toprule
Sector & \textbf{SME count} & \textbf{SME Proportion} & \textbf{No of Employees} & \textbf{Turnover} \\ 
\midrule
Information \&  Communications  Technology & 16 & 18.8\% & 1 - 71 & £10k – £7m   \\ 
Consultancy  & 10 & 11.8\% & 1 - 20 & £10k – £1.4m    \\
Education and Training & 10 & 11.8\% & 1 - 70 & £10k – £6.5m \\ 
Marketing \& Public relations & 7  & 8.2\%  & 1 - 15 & £10k – £250k \\
Human health \&  Social work Activities & 6  & 7.1\%  & 1 - 35 & £10k – £650k \\ 
Travel and Tourism & 5 & 5.9\% & 1 - 5 & £10k – £700k    \\ 
Manufacturing & 4 & 4.7\%  & 1 - 55 & £10k – £850k    \\ 
Accounting \& Finance & 3   & 3.5\%  & 1 - 146 & £10k – £10m     \\ 
Arts, entertainment and recreation      & 3  & 3.5\%  & 1 - 10 & £10k – £400k \\ 
Food & 3 & 3.5\% & 1 - 28 & £10k – £1.2m    \\ 
\bottomrule
\end{tabular}
}
\end{table*}
The top 10 business sectors presented in Table~\ref{tab:Top-10-business-sectors}, together accounts for 78.9\% of the total SMEs from the sample. As expected, most of the SMEs were from the Information \& Communications Technology sector with 16 businesses representing 18.8\% of the total companies involved in the project. Other sectors represented with a higher number of SMEs includes Education and Training (10 SMEs and11.8\% of total), Consultancy (10 SMEs and 11.8\% of total), Marketing \& Public Relations (7 SMEs and 8.2\% of total) and Human health \& Social work activities with 6 SMEs and 7.1\% of total. These numbers suggest that companies in the technology and services sectors rely more on data-driven solutions to support their business when compared to companies in other sectors.

The SMEs included in this analysis completed at least one of the following types of business support (outputs) provided by the BDC project:
\begin{enumerate}
    \item 12 hours business assist and training
    \item Support to introduce a new product/service to the business
    \item Support to introduce a new product/service to the market
\end{enumerate}

Figure~\ref{fig:Count_and_proportion_of_project_outputs_by_type} below shows the count and proportion for each of the three output types. 
\begin{figure}[b!]
\centering
\includegraphics[width=\columnwidth]{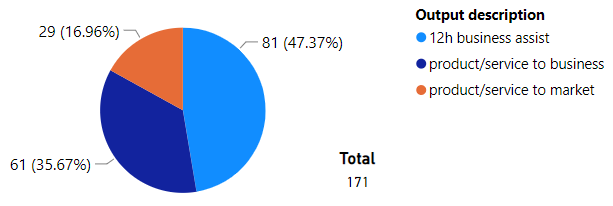}
 \caption{Count and proportion of project outputs by type}
    \label{fig:Count_and_proportion_of_project_outputs_by_type}
\end{figure}
The total number of outputs (171) is higher than the total number of SMEs considering that companies could receive multiple types of business support. More specifically, each company should complete only one 12-hours business assist output, but there was no limitation on the number of new products/services offered to a business and on the new products/services to market outputs. The most prevalent type of output was the 12-hours business assist with a total count of 81 assists, and since each company could only receive one, most of the companies (81 out of 85 or 95\%) benefited from this type of output. This can be attributed to the fact that this type of support included attendance to seminars and workshops with data science related subjects, as well as receiving quick support to solve non-complex data related business problems. From the remaining outputs which required longer-term assistance and collaboration, 61 of the supported SMEs were involved in the introduction of a new product or service to their businesses internally and 29 on the introduction of a new product or service to the market. These results show that companies mostly sought for data-related support to upgrade their internal services, extract insight from their data and improve the decision-making processes. However, we should also mention that some of the outputs that were not introduced to the market were either Proof of Concept (PoC) solutions or components of larger projects aimed to be released to the market in the future.\\

\begin{figure*}[b!]
\centering
\includegraphics[width=.9\textwidth]{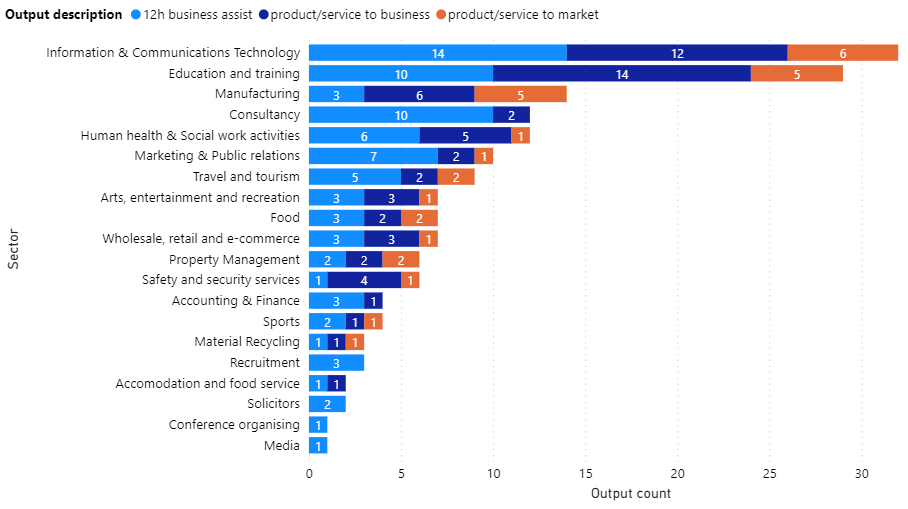}
 \caption{Count of project outputs per category by sector}
    \label{fig:Count_of_project_outputs_per_category_by_sector}
\end{figure*}

Figure~\ref{fig:Count_of_project_outputs_per_category_by_sector} presents the counts per type of output for the twenty different business sectors. 
The sectors with the most outputs were Information \& Communication Technology (32 outputs) and Education \& Training (29 outputs). However, considering the count of SMEs from each sector, the Education \& Training sector had almost 3 outputs per SME (29 outputs for 10 SMEs) while Information \& Communication Technology had 2 per SME (32 outputs for 16 SMEs).In this category, the Manufacturing sector comes first with 14 outputs for 4 SMEs (3.5 outputs per SME). The sector that received most of the 12-hour support was Information \& Communication Technology with 14 outputs followed by Education \& Training and Consultancy sectors with 10 outputs each. The Information \& Communication Technology sector also received most of the product/service to market support with 6 outputs, and Education \& Training sector received most of the product/service to business support with 14 outputs. \\

For the purpose of this analysis, we further introduced four categories of digital support based on the requirements of the support provided to businesses from the project’s academic partners:
\begin{enumerate}
    \item Skills development and training: Companies attended training seminars and workshops focused on data collection, transformation and storage as well as seminars about data analysis, visualisation and using data science and machine learning to get better insight and make data-driven decisions.
    \item Data management and analytics: Supported SMEs to collect and store data more efficiently, in order to make them accessible for analysis and visualisation and allow for better business insight extraction.
    \item System design and development: Supported SMEs to design and develop end-to-end business solutions either to improve their products and services or to enhance internal business decision making and planning.
    \item {Other: This refers to a small number of SMEs provided with a support that does directly fall under one of the above three categories, such as recycling, and fire and safety.}
\end{enumerate}

\begin{figure}[h!]
\centering
\includegraphics[width=\columnwidth]{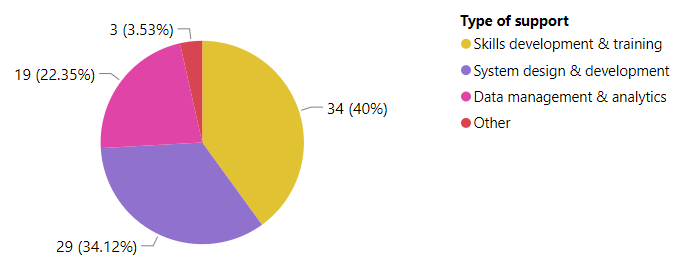}
 \caption{Count and percentage of SMEs per type of support received}
    \label{fig:Count_and_percentage_of_SMEs_per_type_of_support_received}
\end{figure}

As illustrated in Figure~\ref{fig:Count_and_percentage_of_SMEs_per_type_of_support_received}, 34 companies (40\%) required support related to skills development and training, 29 (34.1\%) for system design and development, and 19 (22.3\%) for data management and analytics. Across all different types of support, the main data science and analytics aspects in which the companies required assistance can be summarised as follows:

\begin{enumerate}
    \item Acquiring data (open or proprietary) from external web resources such as APIs, websites or web repositories and transform, filter or combine this data to extract business insight.
    \item Creating advanced visualisations including interactive dashboards and performing descriptive analytics to gain insight and improve products/services or marketing practices.
    \item Collecting, storing and analysing data originating from sensors including IoT.
    \item Using Machine Learning (ML) to improve their products/services with data-driven processes or to provide recommendations for their customers
\end{enumerate}

\begin{figure}[ht!]
\centering
\includegraphics[width=\columnwidth]{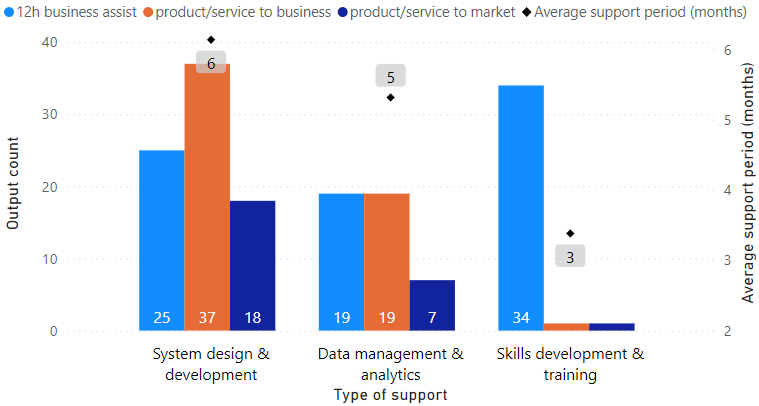}
 \caption{Output count by type and average support period per type of support}
    \label{fig:Output_count_by_type_and_average_support_period_per_type_of_support}
\end{figure}

Figure~\ref{fig:Output_count_by_type_and_average_support_period_per_type_of_support} presents the number of outputs per type of support. It is evident that the project produced higher number of outputs for System Design \& Development and Data Management \& Analytics support when compared to Skills Development \& Training, especially for outputs that assisted companies to introduce new products and services internally or to the market.\\

\begin{figure*}[ht!]
\centering
\includegraphics[width=\textwidth]{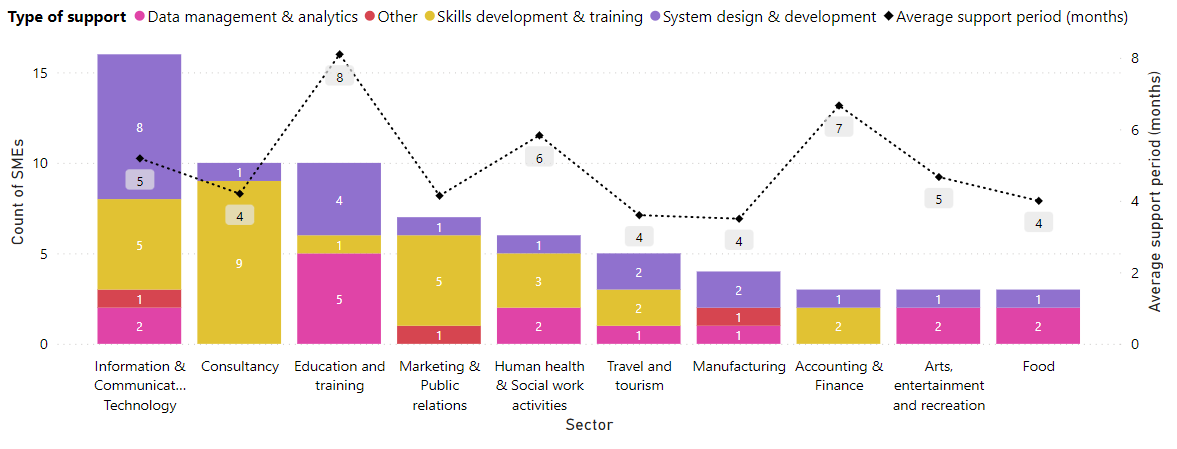}
 \caption{Count of SMEs received support by type of support and average support period per sector}
    \label{fig:Count_of_SMEs_received_support_by_type_of_support_and_average_support_period_per_sector}
\end{figure*}

The duration of support received by SMEs ranges from 1 to 16 months and the average support period was slightly below 5 months (4.75) per SME. As evident in Figure~\ref{fig:Output_count_by_type_and_average_support_period_per_type_of_support}, companies that received support to design and develop data-related solutions or to improve their data management and analytics processes required higher support periods on average (5-6 months) when compared to companies that received support for skills development and training (average support period of 3 months). In Figure~\ref{fig:Count_of_SMEs_received_support_by_type_of_support_and_average_support_period_per_sector} we show the top 10 business sectors supported and the average support period received by SMEs from each sector. Education \& Training sector SMEs had the highest average support period (8 months), followed by Accounting \& Finance (7 months) and Human Health \& Social Work Activities (6 months). In general, it was observed that longer support periods were required for sophisticated projects, which either demanded expert knowledge, or due to decreased levels of data science and technical skills within the companies.

Another important observation derived from Figure~\ref{fig:Count_of_SMEs_received_support_by_type_of_support_and_average_support_period_per_sector} is that System design \& development support was requested by companies across all sectors. This highlights the high demand for data-driven solutions and services in the market and corroborates with the main aim of the study, to encourage digital transformation and use of data in SMEs to achieve growth.

\subsection{Challenges and Learned Lessons}

\begin{figure}[t!]
\centering
\includegraphics[width=\columnwidth]{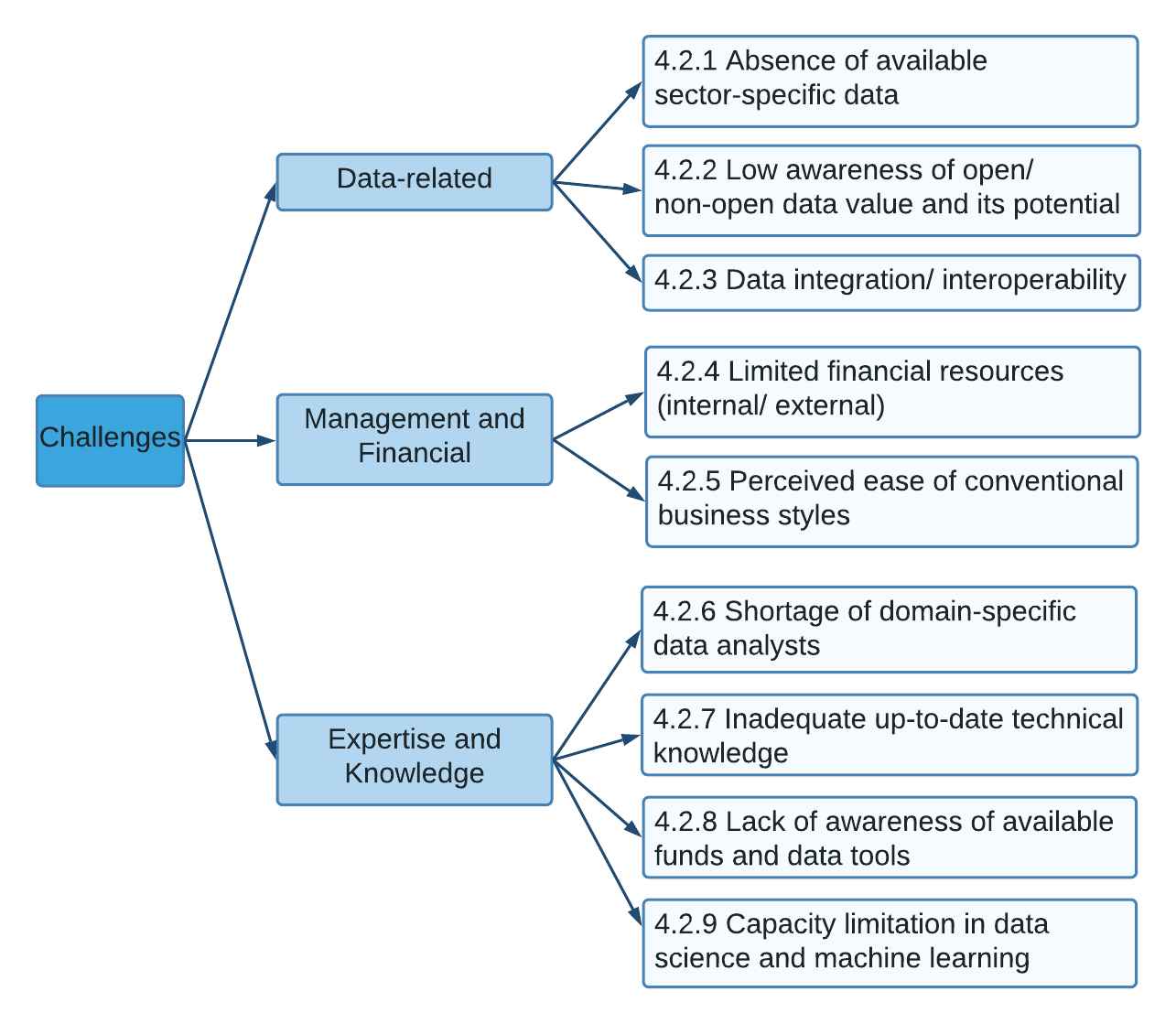}
 \caption{Identified challenges facing SMEs in adopting data-driven approaches}
    \label{SME-challenges}
\end{figure}

The digital transformation of SMEs and their adoption of data science techniques require organizational changes, investment and resource before businesses can reap any resulting profits and benefits. However, we have observed during this study that majority of micro and small enterprises are hesitant to invest in new digital tools and methods if they cannot anticipate quick positive results and revenues. From our analysis, we have empirically recognized   four main    areas (without excluding other potential ones) in which SMEs effectively utilize data-driven strategies and techniques to enhance their business operations. 
\begin{itemize}
  \item First, improving digital marketing proved to be a common data-driven use case among small businesses. Specifically, a large number of the study’s SMEs managed to derive immense marketing insights from analysis performed on their own data or relevant open data, e.g., identifying lead customers, gaining a better understanding of customer purchasing patterns, etc. 
  \item Second, the use of Social Media (SM) platforms for digital marketing was found popular among our case study SMEs. This is driven by the valuable data generated because of customers communicating SMEs via their dedicated channels on these platforms. This SM analytics was used by some enterprises to study the effect of marketing on different platforms (e.g., Instagram, Facebook, LinkedIn, Twitter) on customer purchase and service use intents. This leads SMEs to retain the highest revenue generating platforms and unsubscribe from underperforming ones, eventually reducing costs.  The same analytics was also used to identify working marketing strategies based on specifically studied Key Performance Indicators (KPIs), e.g., views, likes, reach, impressions, etc. Table ~\ref{tab:Top-10-business-sectors} and Figure~\ref{fig:Count_of_SMEs_received_support_by_type_of_support_and_average_support_period_per_sector} show that Marketing and Public relations represents the top support seeking business sectors according to our study. Linked to this, a study on the adoption of social media and SMEs’ performance found that around 70\% of SMEs use social media for marketing~\cite{qalati2021mediated}.
    \item Third,   a good number of the SMEs supported under BDC used   data-driven methods   to   identify    potential new markets   by performing cross-analysis   of their customer data and various relevant open datasets, e.g. government open datasets, such as census, demographics and geospatial data.  Such customer analytics enabled SMEs to locate new markets with a high demand for their services and products, thus allowing them to expand their market presence.
  \item Eventually, performing exploratory analysis on historical data with the objective of creating predictive models was another common application observed among the case study SMEs.  Going to the predictive stage of the analytics ladder enables  many small and medium-sized businesses  to  make use  of  cutting  edge  ML and AI  technologies. The selected two case studies presented Section~\ref{section5} represent good examples for this type of application. They include an SME transforming from manual to digital business operations, and another adopting machine learning for parameter optimization in additive manufacturing. 
\end{itemize}
Linked with the above, the research and consultancy work with SMEs  under this project has also produced other lessons. Especially, we underline a number of factors found to be the leading barriers preventing most SMEs from embracing new and emerging data-driven technologies and AI. Figure~\ref{SME-challenges} summarizes the primary challenges facing SMEs to successfully adopt digitization and data-driven approaches as identified in this study. The challenges are roughly classified into three broad categories: data-related, management and financial, in addition to expertise and knowledge.

\subsubsection{Absence of available sector-specific data}
Some AI and big data analytics applications are relatively new in the context of their deployment and use for some specific domains. This means that any meaningful and successful application in such areas will be hampered by the lack of existing analytic data. For example, in one case of the 85 SMEs in this work, we performed a pilot study of applying supervised machine learning to input-output parameters optimization in Additive Manufacturing. One primary limitation found in this study  was the insufficiency of fabrication data in  the 3D Printing sector. This is not to suggest that there is no reasonable closed data among the SMEs in this sector, but that may not be enough to implement supervised machine learning models and there are no feasible current mechanisms in which such SMEs can share their own hard-earned limited data. This led to suggestions of using simulated data for ML-based parameter optimization in AD~\cite{meng2020machine}.

\subsubsection{Low awareness of open/ non-open data value and its potential}
In many cases, government data set are made available for free and are made public through data openness. Usually, the reasons are related to improving the service, increasing economic value, or improving political transparency. These datasets can be valuable to the success of small businesses, but they seem to be largely unknown to them. Around 20 percent of the SMEs we supported created new services and discovered new markets using open data from education, e-recruitment, and transportation. The more SMEs understand open data's advantages and how it can deliver positive outcomes, the more innovative products and services they are able to design using it at a lower cost. Many of the SMEs we worked with were aware that they could use data not only to keep records but also to gain useful insight into their business. Yet, learning data science skills is a challenge for SMEs. Moreover, they are unsure of the types of data from their operations that should be collected. The fact that SMEs lacked knowledge about data handling tools made them unable to use this data.

\subsubsection{Data integration/ interoperability}
The competitiveness of SMEs can be improved by integrating their systems with their suppliers or other open data available from governments or trading partners ~\cite{mancini2021data}. The demand for businesses to exchange real-time information has grown since the adoption of technologies such as mobile commerce, electronic funds transfers, supply chain management, and online transaction processing. Organizations could benefit from integrating their IT infrastructures to address this need. In many cases, however, portability and interoperability measures for SMEs are still being implemented relatively slowly, while in others they are at an early stage. Data integration is a significant problem for SMEs due to high costs and technological requirements and is frequently cited as being a sticking point within organizations. Some existing approaches to integration, such as electronic data interchange (EDI), inventory management, and automated data collection systems, can help SMEs overcome some of their integration challenges, but they have their limitations. It is possible to further simplify austere integration problems by utilizing web services nowadays. The need for case studies on the integration efforts of SMEs and how they deal with integration and interoperability problems is critical~\cite{themistocleous2004investigating}.

\subsubsection{Limited internal and external financial resources}
In our research work with SMEs, we discovered that most of them are able to recognize and utilize data and analytics to grow their businesses. Also, they we noted their willingness to use data analytics and invest in related technologies to derive meaning from their data. Despite their attempts for adopting data technology, their capabilities are insufficient since they do not possess the required knowledge and expertise, nor the budget to hire specialists or outsource their analytic needs. A further barrier comes from SMEs having limited access to funding and loans compared to big companies. Thus, the significance of the ERDF and other similar projects, which aim to bridge this gap, is justified. A greater level of financial and technical support is also needed to foster the adoption of data analytics by micro- and small-sized businesses, especially during this period of COVID-19~\cite{DigitalTrasformationSMES}.

\subsubsection{Perceived ease of doing business conventionally}
The majority of SMEs operate in specific business sectors and are dependent upon conventional business methods. Thus, many business owners would prefer to maintain their traditional business practices and refrain from adopting disruptive technologies, such as artificial intelligence, data analytics, etc. As a consequence, they are not tempted to use data other than for record-keeping purposes. There are a number of possible explanations for this reluctance to leverage the benefits of advanced data usage-not just a misunderstanding of what these advances are meant to offer their companies, but also the perceived short-term disruptions such changes might cause-such as learning new tools, implementing software packages or investing in cloud computing, and hiring employees with data analytics skills. However, the result is an overlooked data-driven business opportunity, which could have been beneficial for the SMEs in the long run. SMEs' decision-making process could be made more data-driven by encouraging a business shift.

\subsubsection{Shortage of domain-specific data analysts}
It is often necessary to combine data analytics skills with business context and domain knowledge in order to analyze business data effectively~\cite{banerjee2013dat}. As a result of the limited number of analysts in the business who meet such criteria, SMEs outsource their data analytics needs to experts at comparatively higher costs. Data science and machine learning for process control and modelling are still in their infancy in some domains, e.g., additive manufacturing, due to the lack of many experts who understand the business context. SMEs in manufacturing are less likely to use data for analytics and decision-making~\cite{soroka2017big}, which makes it harder for them to exploit their data to enhance growth and productivity.
\subsubsection{Lack of or outdated technical knowledge}
Small businesses need to develop and upgrade employee skills in order to grow. The lack of finance, the lack of capacity in data science and machine learning, and the lack of awareness of innovation funds, among others, prevent many SMEs from remaining competitive as technology advances in terms of employee skills and competence. These are especially applicable to smaller companies, such as manufacturing and technology companies, who face challenges when designing, developing, and testing new products or upgrading their existing ones to meet new needs. Based on Section~\ref{SMEsDigitizationTrends} analysis, we can understand that the percentage of SMEs with skills and training development needs is roughly  40\% as shown in Figure~\ref{fig:Count_and_percentage_of_SMEs_per_type_of_support_received}.

\subsubsection{Insufficient knowledge of the data tools and funds available}
In our study, we found that SMEs are unaware of the available funding, especially for adopting innovative digital and data-driven ideas. Knowledge Transfer Partnerships (KTP) are considered one of the most appealing funding schemes with respect to the BDC's digital assistance and collaboration with SME business owners since minimal contributions are required from SMEs and the partnership may have a high level of impact on SMEs. In addition, our study found that only a very small percentage of micro enterprises are aware of some of the most widely used data management and analysis tools, and the benefits these tools can provide. It proved crucial to many SMEs to support their adoption of these tools (including Dropbox, Power BI, and R Studio).

\subsubsection{Capacity limitation in data science and machine learning}
One other primary challenge facing SMEs in digitizing and adopting AI and data science technologies is the inadequacy or lack of working knowledge and capacity in these technologies. This barrier is partially linked to the other aforementioned challenges such as SMEs limited resources, e.g. finance. Figure~\ref{fig:Count_and_percentage_of_SMEs_per_type_of_support_received} shows the breakdown analysis of the digital technology support sought by the sample SMEs of which \emph{skills development and training} forms the largest percentage (40\%) followed by \emph{system design and development} (34\%). For some SMEs, this was partially compensated by outsourcing data science and AI  solutions from external expertise such as cloud computing-based Machine learning as a Service (MLaaS) platforms~\cite{DigitalTrasformationSMES}.

%{\bf A Taxonomy of challenges (barriers)}

\section{Case studies in digitization and adoption of AI and analytics aspects in SMEs }\label{section5}
In this section, we present two illustrative case studies. The first one (Section~\ref{PBL_Care}) is about an SME that transformed from manual business operations to digital processing under this research and support project. The second case study looks at an additive manufacturing company that embarks the route to  AI and Machine Learning adoption for parameter optimization (Section~\ref{HiETA})
% for tables, first create it using https://www.tablesgenerator.com/latex_tables
% then replace tabular by tabularx
% finally, add "{\textwidth}{|A{0.4}|A{1.6}|}" after \begin{tabularx} with relevant values of column ratio (adding up to the total number of columns, 2 in the example)

\subsection{A data-driven solution for monitoring the delivery of PBL care services}\label{PBL_Care}
PBL Care is a domiciliary care service SME based in Birmingham - West Midlands \cite{PBLCare}. It provides home care and support to individuals who live independently in their own homes. The SME offers a wide range of services such as personal care, assistance with eating and toileting, medication support, and palliative care. Among the individuals supported by PBL Care, some have physical disabilities, dementia, and mental health conditions. PBL Care is regulated by the Care Quality Commission (CQC), part of the UK Department of Health and Social Care, responsible for the regulation and inspection of health and social care services in England \cite{CQC}.

\subsubsection{Problem statement}
The CQC inspected PBL Care on December 2017 and rated the SME as “Requiring improvement”, meaning that the service is not performing as well as it should. This rating is the 3rd of 4 possible rating given by the CQC, the 4th one being "Inadequate". The CQC report highlighted that at the time of the inspection, the company did not have the right processes in place to guarantee an effective monitoring of the delivery of care, resulting in late arrival of care staff at patient's homes and most visits not lasting as long as planned. In addition to this, most of the data collected from visits were paper based. The literature describes that digitisation of patient records can improve communication and coordination in health care organisations~\cite{atasoy2019digitization, mihailescu2018emergence}, especially in home-care context~\cite{soikkeli2013evaluating}. Accordingly, following several meetings with the SME to understand requirements, it was agreed to digitise the business in order to meet evolving demands and keep pace with the rapidly changing home care services. 

\subsubsection{Methodology}
The case study aims at creating a data-driven solution for PBL Care that reflects the implementation of new processes and allows the SME to monitor delivery of care digitally. As a result and following a recent inspection, the CQC provided the following guidance to the business: "your care services, treatment and support achieves good outcomes, helps you to maintain quality of life and is based on the best available evidence.".

We list below the workflow steps of our collaboration plan with PBL Care for its digitization: step 1: Identification of Key Performance Indicators (KPIs); step 2: Data collection; step 3: Data visualisation and dashboard prototyping; and step 4: Data interpretation and evaluation. The aim of the last step is to analyse outcomes following the implementation of new processes by PBL Care management team; an analysis that could lead to further process changes. In the following sections we will describe each of these steps.

\subsubsection {\textbf{}Development of a digital solution}
\paragraph{\textbf{Identification of KPIs}}
In order to identify KPIs, we started by looking at paper-based information collected by PBL Care. The SME used paper ``log books" to record information such as date, time in, time out, carer's name, detailed log of tasks carried out and any concerns raised, food/fluids taken, pad changed, etc. Care staff fill one log book per visit at the service user (patient) location. We identified the following attributes as key for monitoring: \textit{Patient ID, Care staff ID, Date, Time In, Time Out}. 

\paragraph{\textbf{Data Collection}}
Data collection templates were created using Excel spreadsheets to digitally gather data based on the identified KPIs. For the first iteration of data collection and dashboard prototyping, care staff continued to fill in paper log books at the patients’ homes and other staff inputted these data in the spreadsheet at PBL Care office. We will also discuss how this process was later improved. 

% Please add the following required packages to your document preamble:
% \usepackage{booktabs}

\begin{figure*}[b!]
\centering{
\includegraphics[width=\textwidth]{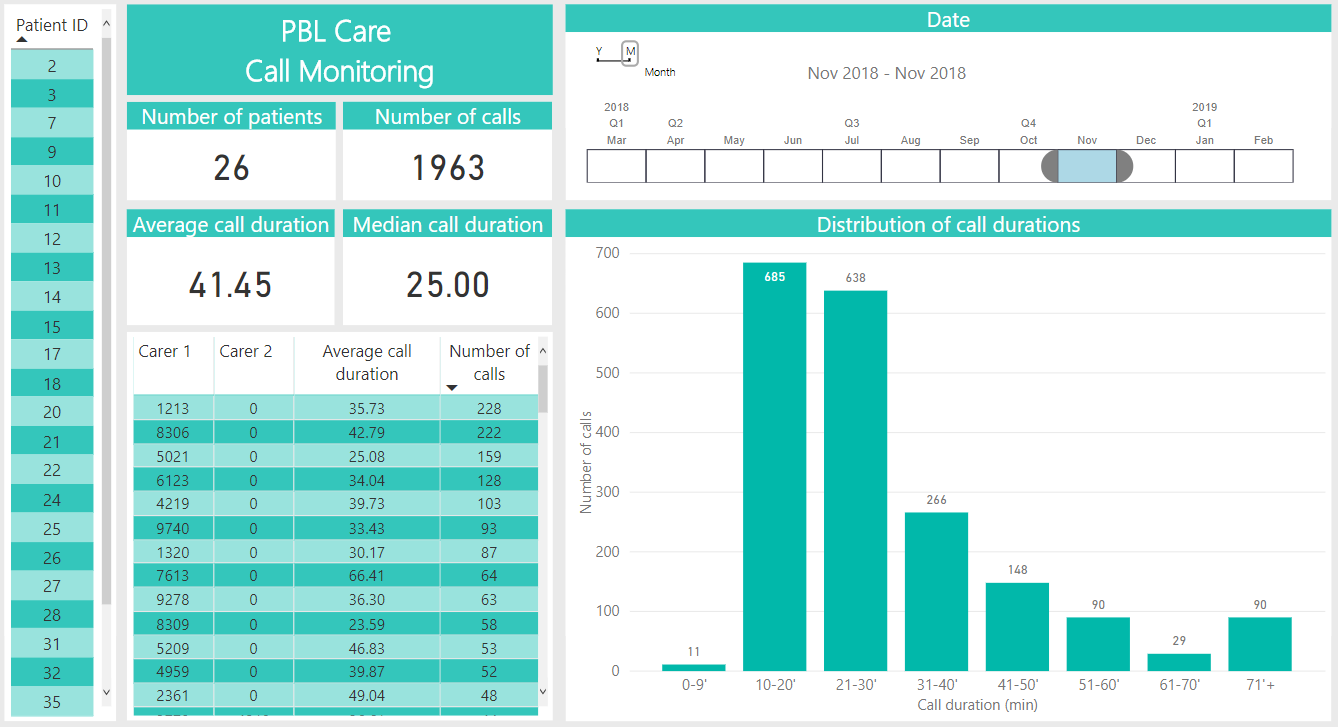}}
 \caption{Dashboard prototype for monitoring of visits duration}
 \label{fig:MainDashboard}
\end{figure*}

\paragraph{\textbf{Data visualisation and dashboard prototyping}}
The National Institute for Health and Care Excellence (NICE) guidance states that “home care visits to elderly people should last for at least half an hour [...], visits shorter than half an hour should only be made under specific circumstances". Given this, we built an interactive dashboard to monitor this KPI "visit duration" using Microsoft PowerBI Business Intelligence (BI) tool (Version: 2.6) as presented in Figure~\ref{fig:MainDashboard}. The dashboard allow for users to select a month, a service user (patient), a care staff, or a histogram range by clicking on an id or time range and as a result the rest of the dashboard is filtered based on the selection. The dashboard  allow for the PBL Care management team to look specific service users or care staff, and also to check whether NICE guidance on visit duration is followed. A click on “refresh” in the dashboard pulls the latest data entered in the spreadsheet, making it easy for PBL Care to visualise the latest data. Additional dashboard pages on other topics covering food served to service users, types of home care services provided, targets set for patients, and spot check information were also built to support he SME.

\paragraph{\textbf{Data interpretation and evaluation}}
From the histogram of Figure~\ref{fig:MainDashboard}, we observe that most calls normally last for a duration between 10 and 20 minutes (685 calls), which is not a good practice as mentioned earlier. This might be due to calls scheduled back-to-back, not allowing enough time for travel between visits and pushing care staff to hurry and rush with providing care. PBL Care took some measures to address this: the care staff scheduled visits were adjusted to allow sufficient travel time between patient homes; also, the care staff were reminded of the importance to be on time for their visits and made aware that ``spot checks" would be carried out.

\begin{figure}[ht!]
\centering{
\includegraphics[width=\columnwidth]{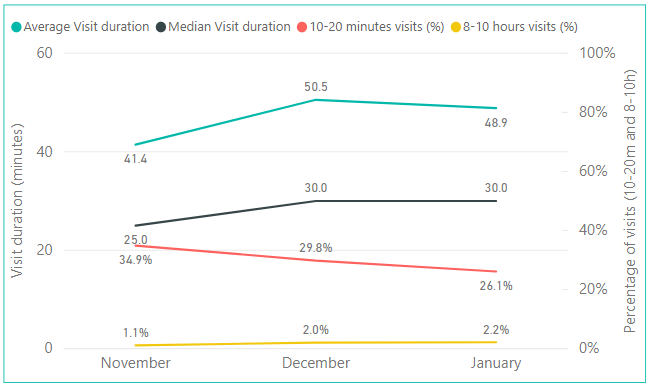}}
 \caption{KPI evolution}
 \label{fig:KPIevolution}
\end{figure}

The rest of our interpretation is focused on the analysis of three consecutive months (Nov 2018, Dec 2018, and Jan 2019) for which data were collected consistently over time and for a similar number of patients (respectively 26, 36, and 31). Figure~\ref{fig:KPIevolution} presents the average visit duration, median visit duration in minutes, percentage of visits between 10 and 20 minutes, and percentage of visits between 8 and 10 hours for the months listed earlier.

The average visit duration is around 50 minutes for December and January, with an increase of about 8 minutes from November. This is partly because PBL Care took more night-shift NHS packages from December, as shown by the increase in percentages of visits between 8 to 10 hours from 1.1\% in November to 2.0\% in December. The median visit duration, more robust with regards to extreme values, is 30 minutes in both December and January. We can observe a consistent decrease in the percentage of visit duration between 10 and 20 minutes, from 34.9\% in November to 26.1\% in January. This decrease could be linked to the two actions mentioned earlier that PBL Care management team took (care staff schedules and spot checks). \\

The dashboard prototyping and preliminary analysis have shown to PBL Care the merit and capability of our proposed digitization approach in monitoring KPIs of SME's interest over time through interactive dashboards with the aim to improve the quality of services provided to patients. \\
\subsubsection{Findings - Improving provision of care through data reporting}
Following the creation of the dashboard prototype, PBL Care decided to use it during monthly meetings with care staff to show progress and to set actions. Based on the collected data, dashboard visualisations can highlight problems in an organisation and lead to the modification and implementation of new processes. In an approach of continual improvement process, these steps can then be repeated and as the number of iterations grows, it is likely to observe improvements in the services provided and in the solution developed.
% as illustrated in Figure~\ref{fig:ContinualImprovement}

%\begin{figure}[t!]
%\centering{
%\includegraphics[width=\columnwidth]{Images/cycle-schema-Figure-B.PN%G}}
% \caption{Continual Improvement Process}
% \label{fig:ContinualImprovement}
%\end{figure}

Conscious of the importance of collecting data digitally, PBL Care started to use an accredited software provider from February 2019, including a mobile application for care staff and a web application for PBL Care management team. In March 2019, the Care Quality Commission conducted a new inspection of PBL Care which resulted in a better rating from “Requiring improvement” from "satisfactory" to “Good”.\\
We have learnt that a digital solution can help monitor the provision of care. However, it must not be the only source of information for driving decisions. Underlying factors can be at play behind the scenes and not be reflected on the dashboard. The digital tool can help generating hypotheses and highlight patterns that need to be raised in conversations between management staff, care staff, and service users. For example, the management team must not jump to conclusions and blame a care staff if the dashboard shows that an employee had a couple of visits that lasted 5 minutes. A simple reason like a service user telling the carer he did not require care on that day could be the explanation. The identification of KPIs and interpretation of data must not be based only on performance, it must also reflect the provision of good care and satisfaction of individuals, both service users and care staff.

\subsection{Process parameter optimization for Additive Manufacturing using supervised machine learning}\label{HiETA}
Additive  Manufacturing  (AM) is the  process  of  fabricating  components  by  adding  layer-upon-layer  of materials with the aid of digital 3D design~\cite{delli2018automated}.  AM has recently gained increasing research attention because of its advantages in comparison with traditional subtractive manufacturing~\cite{qi2019applying}. Although, the use of machine learning for additive manufacturing is still in its infancy stage, several machine-learning algorithms have been applied in AM tasks including parameter optimization and fault detection~\cite{meng2020machine,delli2018automated}. Related to this, Meng et  al.~\cite{meng2020machine} reviews  latest  applications  of  different  machine  learning  algorithms  in  additive manufacturing. The study matches various ML methods to corresponding AM applications including parameter  optimization  and  anomaly  detection. 

This  case  study  is about   parameter  optimization  in 3D printing for a company called HiETA Technologies\footnote{https://www.hieta.biz/}. It attempts to investigate a model for establishing relationship between specified process inputs and defect indicators in produced components. HiETA Technologies is a research and product development company founded in 2011. The SME specializes in the use of Additive Manufacturing (metal 3D printing) for thermal management and light-weighting solutions. They offer  an end-to-end metal  3D printing services  in  additive manufacturing and  engage in  development  projects  for  a  range  of  energy  systems  including fuel  cells, turbine machinery, nuclear,  concentrated  solar  power, and  other heat  and  power  generation  systems  for automotive and airspace applications. HiETA’s unique technical capabilities and technologies include the development of heat transfer surfaces with increased high heat transfer. The company strives to create methods and processes that dramatically reduce product size, and increase cycle efficiencies and product life. 

\subsubsection{Problem statement}
HiETA Technologies currently uses a series of trial and error testing (printing component samples repeatedly and  testing  their  quality  and  defects)  to optimize process  parameters  and  get  to  optimal input values for the production of final components. This brute force method, commonly used in the wider 3D printing business sector, does not only cost significant amount of time causing major delays, but also incurs countless failures and mistakes before yielding the right parameters that work for the given production. To address this issue, the company wants to investigate the possibility of using machine learning in their processes to reduce errors and the high time and material costs associated with the use of their existing trial and error method.  On  that  initiative,  HiETA  joined  the  BDC  in a research collaboration to investigate this problem.

\subsubsection{Initial data preparation and exploration}
HiETA’s primary objective of applying machine learning in its additive manufacturing process is to automatically identify the optimal parameters for building components with few or no trails.  In  particular,  the task of this first stage is to   develop a bespoke predictive  model to be used in the material characterisation of new product development. This will in the long term  optimise  process  parameters  for  given  geometry-material-machine  combination,  thus improving the SME’s understanding of parameter interactions and cause-effect relationships in the AM process. Such predictive model will also establish a relationship formula between the input and target parameters enabling HiETA to run the minimum possible experiments to get the right parameter values that work. This will also focus production time and resources  into  a small  pool  of  experimental  trials. 
  
  HiETA provided the BDC project with sample data of six parameters for  this  pilot investigation. Table \ref{tab:HiETA-statistics}  shows anonymised statistical summary  of the sample  parameters data and  their modelling roles. \\
    % Please add the following required packages to your document preamble:
% \usepackage{booktabs}
\begin{table*}[b!]
\centering
\caption{Summary statistics of the Sample data parameters and roles (rounded to 3 S.F)}
\label{tab:HiETA-statistics}
\begin{tabular}{l@{\hskip 0.5in}  l@{\hskip 0.5in}  l@{\hskip 0.5in} l@{\hskip 0.5in} l@{\hskip 0.5in} l@{\hskip 0.5in} }
%\begin{tabularx}{0.95\textwidth}{l@{\hskip 0.5in} c@{\hskip 0.5in} c@{\hskip 0.5in} c@{\hskip 0.2in}c@{\hskip 0.5in} c@{\hskip 0.2in}}
\toprule
Parameter & Mean  & Std deviation & Median & Skewness & Role  \\ [0.5ex] % inserts table 
\hline % inserts single horizontal line \\ %\midrule
ET        & 0.645 & 0.800         & 0.645  & 0.000    & Input (predictor) \\ %\midrule
HO        & 0.387 & 0.533         & 0.387  & 0.000    & Input (predictor) \\ %\midrule
LP        & 1.806 & 1.600         & 1.806  & 0.000    & Input (predictor) \\ %\midrule
PD        & 1.161 & 1.067         & 1.161  & 0.000    & Input (predictor) \\ %\midrule
ABULD     & 0.752 & 0.583         & 0.851  & 0.638    & Output(excluded)  \\ %\midrule
ALD       & 1.248 & 1.339         & 1.167  & 1.163    & Output(excluded)  \\ %\midrule
S1BLD     & 1.359 & 1.698         & 1.185  & 1.237    & Output(response)  \\ %\midrule
S1BULD    & 0.429 & 0.430         & 0.394  & 0.619    & Output(excluded)  \\ %\midrule
S2BLD     & 1.313 & 1.362         & 1.184  & 1.175    & Output(response)  \\ %\midrule
S2BULD    & 1.149 & 0.844         & 1.436  & 0.540    & Output(excluded)  \\ %\midrule
S3BLD     & 1.073 & 1.094         & 1.045  & 1.194    & Output(excluded)  \\ %\midrule
S3BULD    & 0.677 & 0.651         & 0.738  & 1.435    & Output(excluded)  \\ 
\bottomrule
\end{tabular}
\end{table*}
You will note that four of data parameters, namely; laser power (LP), point distance (PD), hatch option  (HO),  and  exposer  time  (ET)  are  predictors  whereas  border  length  densities  (S1BLD, S2BLD, S3BLD)  and bulk  length densities  (S1BULD,  S2BULD, S3BULD) for  samples 1,  2,  and 3    are  response variables.  The  table  shows  that all  four  input  parameters  are  normally  distributed  as  shown  in  their mean-median  equality  and  their skewness.  This suggests  that there  is  no  need for  performing  any variable transformation prior to training a predictive model.  The roles indicated by \emph{output (excluded)} are not included in the modeling exercise as these were not a priority for the SME at  the time of this research collaboration. Thus, the four predictors (LP, PD, HO, ET) and the two SME prioritised target variables (S1BLD and S2BLD) are extracted from the sample data to investigate a predictive model for minimising the defect densities in samples 1 and 2.

\begin{figure*}[b!]
\centering
\includegraphics[scale=0.7]{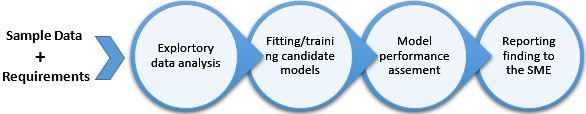}
 \caption{Parameter Modeling Workflow}
    \label{fig:ParameterModelingWorkflow}
\end{figure*}

Figure~\ref{fig:ParameterModelingWorkflow} summarizes the parameters workflow we used to investigate the development of  the  predictive  model. Data  exploration  was  the  first  step  of  the  analytic  workflow, which examined data characteristics including the importance and correlations of input variables to output parameters. We have also looked at the importance of the four process inputs to the two target variables. 

\begin{figure}[t!]
\centering
\includegraphics[scale=0.85]{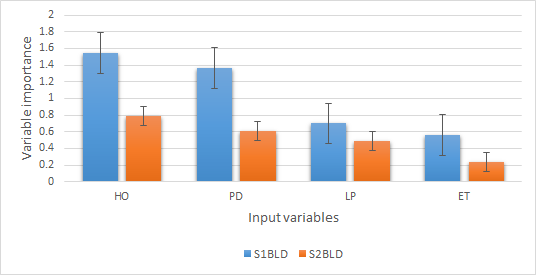}
 \caption{Importance of process inputs for the prediction of output parameters(targets)}
    \label{fig:process-inputs-for-the-prediction-of-output-parameters}
\end{figure}

Figure~\ref{fig:process-inputs-for-the-prediction-of-output-parameters} (cf. Table ~\ref{tab:input-variable-correlations}) shows the linear correlations between the four process input parameters in the sample dataset and the target variables.  It is evident that all four input parameters are better predictors for   sample 1 border length density (S1BLD) compared to that of sample 2 (S2BLD). It is also clear that  the  HO  and  PD  process  inputs  are  more  important  than  LD  and  ET  in  predicting  both defect densities, S1BLD \& S2BLD. Overall,  the low linear correlation r-values given in Table ~\ref{tab:input-variable-correlations} suggest  that  nonlinear  machine  learning  models may better predict the product defects than their linear counterparts, as we will see later.

% Please add the following required packages to your document preamble:
% \usepackage{booktabs}
\begin{table}[b!]
\centering
\caption{Input variable correlations with response variables}
\label{tab:input-variable-correlations}
\begin{tabular}{l@{\hskip 0.3in} l@{\hskip 0.3in} l @{\hskip 0.3in}}
 \toprule
Input variables & \multicolumn{2}{l}{Target variables} \\ \cline{2-3}
   & S1BLD              & S2BLD              \\ \hline 
Exposer  Time(ET)              & 0.13798            & 0.12041            \\ 
Laser Power(LP)              & 0.09345            & 0.11580            \\ 
Point Distance(PD)              & -0.19764           & -0.07931           \\ 
Hatch Option (HO)              & -0.41878           & -0.39791           \\
 \bottomrule
\end{tabular}
\end{table}

\subsubsection{Modeling process parameters using regression algorithms}
The target variables of interest in this predictive modeling task are continuous-valued. This makes regression models the most appropriate predictive modeling algorithms to be used. To this end, we have applied six different supervised algorithms to the sample data, namely:  linear, polynomial, decision tree, random forest, SVM and Multilayer perceptron regression methods. Regression algorithms are  supervised  machine  learning algorithms. This section presents the application of these six regression algorithms to the sample process input-output parameters data.

This category of algorithms require a sufficient amount of training data to produce good  and reliable  predictive  models.  One challenge  facing HiETA,  and  perhaps  the wider AM research and product development companies, is the lack or insufficiency of past fabrication  data,  which  led for some  researchers to suggest  the  use  of  simulated data  for ML-based parameter optimization~\cite{meng2020machine}. Insufficient training data is a common challenge in machine learning, particularly for domains starting to adopt AI or in which the process of creating the training data is very expensive and time consuming such as additive manufacturing~\cite{meng2020machine, lateh2017handling}. To mitigate that, we have used SMOGN algorithm~\cite{branco2017smogn} to oversample the data into various sizes and up-to 2000 instances as the original data was just less than 100 observations. It is widely accepted in AI/ML that predictive accuracy improves with increased training data ~\cite{banko2001scaling, halevy2009unreasonable}. Resampling limited training data and synthesising new data instances is a common practice in predictive machine learning with the objective of increasing small-sized datasets. 

To build a product defect predictive model, we have applied 5 regression algorithms to the oversampled data of up-to 2000 instances. Tables \ref{Results:500}-\ref{Results:1000} summarize the Root Mean Square Error (RMSE) results of  the models built with these algorithms and  the best over-sampling rates, 500 and 1000. The RMSE is   a   regression performance measure used to evaluate the differences between actual and model predicted parameter values. It is computed as per Equation~\ref{RMSE}, where $n, y_{i}, \hat{y}_{i}$ represent the sample size, actual and predicted values respectively. 

\begin{equation} \label{RMSE}
RMSE = \sqrt[]{\sum\limits_{i=1}^n\frac{(\hat{y}_{i}-y_{i})^2}{n}}
\end{equation}

Tables~\ref{Results:500}-\ref{Results:1000} shows that decision trees, random forest, and polynomial models perform better for this sample data compared to other applied regression models. Polynomial models are known for their high computational complexity with increasing data variables and degrees. Overall, it is clear from this set of results, that decision trees produce lowest prediction error consistently across all models and for all oversampling rates\footnote{This includes results achieved with other over-sampled data sizes above 1000 and up-to 2000 instances.}. The fact that linear regression has the lowest accuracy among all models suggests that the relationship between manufacturing defects and input parameters are nonlinear. This corroborates the findings in the initial data exploration, e.g.,  the correlation coefficients between the predictors and target parameters as in Table~\ref{tab:input-variable-correlations}

\begin{table}[ht!]
%\begin{tabularx}{0.95\textwidth}{l@{\hskip 0.1in} l@{\hskip 0.1in} l@{\hskip 0.1in}
 		%\captionsetup{width=0.9\textwidth}
\caption{Defect Prediction performance of different regression algorithms based on oversampled data of size 500 instances: all figures are rounded to 3 S.F}
 		
 		\label{Results:500}
\begin{tabular}{l@{\hskip 0.13in} l@{\hskip 0.12in} l@{\hskip 0.12in}l@{\hskip 0.12in} l@{\hskip 0.12in}}
%\begin{tabular}{l@{\hskip 0.3in} l@{\hskip 0.3in} l @{\hskip 0.3in}}
 		%\begin{tabularx}{| l@{\hskip 0.5in} | l | l | l | l |} 
        \toprule
 		%\backslashbox{\textbf{Measure}}{\textbf{Metric}}
 		{\textbf{Algorithm}}&{\textbf{S1BLD}}&{\textbf{S2BLD}}&{\textbf{S1BULD}} & {\textbf{S2BULD}}\\
 		\midrule
 		Baseline  &  1.475 &	1.046&	0.474&	0.865 \\
 		Linear   & 1.32	& 0.926	& 0.384	& 0.74 \\
 		Polynomial  & 0.485	& 0.259	& 0.117 & 0.231 \\
 		\textbf{Decision Tree}   &  \textbf{0.410}	& \textbf{0.25} &	\textbf{0.090} &	\textbf{0.126} \\
 		Random Forest  &  0.461	& 0.289 &	0.104 &	0.161\\
 		SVM (rbf)  &  0.777	& 0.423 &	0.139 &	0.303 \\
 		Neural Network  & 0.598	& 0.314	& 0.140 &	0.251 \\
 		\bottomrule
 \end{tabular}
\end{table}

\begin{table}[ht!]
\caption{Defect Prediction performance of different regression algorithms based on oversampled data of size 1000 instances: all figures are rounded to 3 S.F}
 
 \label{Results:1000}
\begin{tabular}{l@{\hskip 0.13in} l@{\hskip 0.12in} l@{\hskip 0.12in} l@{\hskip 0.12in} l@{\hskip 0.12in}}
 		%\begin{tabularx}{| l@{\hskip 0.5in} | l | l | l | l |} 
 		%\captionsetup{width=0.9\textwidth}
 	     \toprule
 		%\backslashbox{\textbf{Measure}}{\textbf{Metric}}
 		{\textbf{Algorithm}}&{\textbf{S1BLD}}&{\textbf{S2BLD}}&{\textbf{S1BULD}} & {\textbf{S2BULD}}\\ 
 		\midrule
 		Baseline  &  1.718	& 1.188 &	0.513 &	0.954\\
 		Linear   & 1.282&	1.025 &	0.380 &	0.786 \\
 		Polynomial  & 0.433	& 0.293 &	0.144&	0.286\\
 		\textbf{Decision Tree}   &  \textbf{0.379}	& \textbf{0.257} &	\textbf{0.092} &	\textbf{0.114} \\
 		Random Forest  & 0.404 & 0.284 &	0.106 &	0.140 \\
 	SVM (rbf)  & 0.545 &	0.363 &	0.148 &	0.335 \\

 		Neural Network &  0.466 &	0.753 &	0.165	& 0.285 \\
 		\bottomrule
 	\end{tabular}
 	\end{table}
 In Table~\ref{Results:average}, we show the average performance of all algorithms across varying over-sampled data sizes including original one~\footnote{The size of the original data could not be disclosed due to SME's business confidentiality}.
 
\begin{figure}[ht!]
\centering
\includegraphics[scale=0.65]{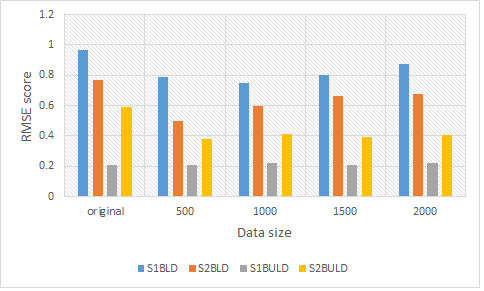}
 \caption{Oversampled data size effect on model performance}
    \label{fig:datasizeffect}
\end{figure}

 Figure~\ref{fig:datasizeffect} visually illustrates the data size effect on prediction performance as presented in Table~\ref{Results:average}. Empirically speaking, the oversampled data in the range between 500-1000 produces best overall average results across models. In other words, the average prediction errors produced with over-sampled datasets 1500 and 2000  for nearly all models are higher than those achieved  with lower level oversampling, e.g. the 500 and 100 observations. This may be due to the fact that the noise in the data gets significantly amplified at the higher oversampling rates. 
 
\begin{table}[b!]
\caption{Average dataset performance over all regression algorithms algorithms }
 		
 		\label{Results:average}
\begin{tabular}{l@{\hskip 0.2in} l@{\hskip 0.2in} l@{\hskip 0.2in} l@{\hskip 0.2in} l@{\hskip 0.2in}}
 		%\begin{tabularx}{| l@{\hskip 0.5in} | l | l | l | l |} 
 		%\backslashbox{\textbf{Measure}}{\textbf{Metric}}
 		\toprule
 		{\textbf{Data size }}&{\textbf{S1BLD}}&{\textbf{S2BLD}}&{\textbf{S1BULD}} & {\textbf{S2BULD}}\\ 
 		\midrule
 		original & 0.965 &	0.770 &	0.206 &	0.590 \\
 		500  &  0.790&	0.501 & 	0.207 &	0.383 \\
 		1000   & 0.747 & 0.595 &	0.221 &	0.414 \\
 		1500  & 0.802 &	0.664 &	0.207 &	0.391 \\
 		2000  & 0.872&	0.680&	0.221 &	0.403 \\
 		\bottomrule
 	\end{tabular}
 		\end{table}

 \begin{table}[t!]
 \caption{Decision tree models results for S1BLD target with varying maximum depth and over-sampled data sizes }

 \label{Results:DT_vary_max_depth}
 \begin{tabular}{l@{\hskip 0.3in} l@{\hskip 0.25in} l@{\hskip 0.25in} l@{\hskip 0.25in} l@{\hskip 0.25in}}
        \toprule
 		%\begin{tabularx}{| l@{\hskip 0.5in} | l | l | l | l |} 
 		%\backslashbox{\textbf{Measure}}{\textbf{Metric}}
 		{\textbf{Tree depth }}&{\textbf{500}}&{\textbf{1000}}&{\textbf{1500}} & {\textbf{2000}}\\ 
 		\midrule
 		3 &	0.779	& 0.711 &	0.768 &	0.857 \\
        6 &	0.512	& 0.455 &	0.442 &	0.500 \\
        9	& 0.446 &	0.400	& 0.378 &	0.419 \\
        12 &	0.420 &	0.383 &	0.355 &	0.387 \\
        15 &	0.415 &	0.379 &	0.346 &	0.377 \\
        18 &	0.411 &	0.379 &	0.343 &	0.373 \\
        21	& 0.410 &	0.379 &	0.343 &	0.372 \\
        24 &	0.410 &	0.379 &	0.343 &	0.371 \\
        27 &	0.410 &	0.379 &	0.343 &	0.371 \\
        30 &	0.410 &	0.379 &	0.343 &	0.371 \\
 		\bottomrule
 	\end{tabular}
 \end{table}
 
 	\begin{figure}[t!]
\centering
\includegraphics[scale=0.55]{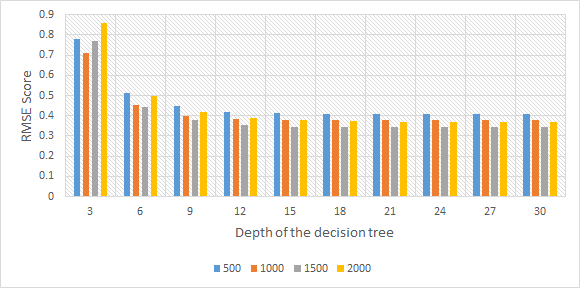}
 \caption{Effect of tuning decision tree depth on model performance}
    \label{fig:varying_depths}
\end{figure}

To investigate any possible further improvement that might be achieved with the best identified model, we have tuned the decision tree depth hyper-parameter. For illustration purpose only Table~\ref{Results:DT_vary_max_depth} and Figure~\ref{fig:varying_depths} show the model errors (hence accuracy) for the S1BLD target variable for different oversampled training data sizes while varying decision tree depths. The results show the model accuracy improves with increasing decision tree depth, although the errors become stable from a maximum depth of 21. In addition, the over sampled data of size 1500 instances consistently achieves the lowest error with various models. 

Based on our analysis, there are several reasons for the weakness of tested models with the given sample data size ($n<100$) been the primary one. The analytic sample data was also found to be noisy (confirmed by the SME), which also contributed to the high model errors.  Statistically  speaking,  it  is widely  accepted  in  the  literature to expect a high error margin when using small sample data with a statistical model, particularly with multiple predictors~\cite{kroll2013impact}. Overall, with less noisy data in place, we think this will be a good step in paving  the  way towards  a  full-fledged parameter  optimization predictive tool to be used within HiETA’s AD process and new  product  development. It  was  also  encouraging  to learn that another  independent (commercial) investigation of the  sample data concluded similar findings based on the SME feedback to us.

%\section{Recommendations}\label{section?}
%\import{sections/}{recommendations}

%\section{Conclusion and Future Work}\label{section6}
\section{Summary and Conclusion}\label{section6}
Digitalization in SMEs has never been more needed than in this COVID-19 era as most businesses adopted online and hybrid operations with many of their employees working from home. In fact, earlier studies in the pandemic have found that up to 70\% of SMEs stepped up digital technology use during the COVID-19~\cite{oecd2021digital}. However, considering constraints including such as the lack of sufficient finance to invest in IT infrastructure and to hire the right skilled experts, SMEs are still far from being in full swing as part of the digital revolution.

Evidently, there is a need to raise awareness among SME owners, managers and entrepreneurs about the advantages and challenges digitalisation could bring to their business, and how different subfields of data science could apply to different industries, business functions and business models. Decision makers have to train in order to rethink their business processes and to reconfigure tasks and organisational structures. More staff would also require upskilling in order to consider and guide analytical outputs, and take data-driven approach to solve problems leading to more informed decisions that benefit the business. The range of challenges identified as a result of our analysis includes:
\begin{itemize}
\item need to Support SMEs in building a culture of data, from collection, to management, to protection and processing, in order to ensure that the digitisation transition takes place with the least risk to SMEs.
\item raising awareness on the benefits of data science and analytics to the business.
\item upskill SME managers and employees and ensuring an involved approach for redesigning business processes and training required to run applications and analysing results.
\item consider mechanisms to bridge the financing gap until the data science solution can deliver its full potentials.
\item enable SMEs to gradually increase their capacity before being eventually able to develop their own data science solutions.
\item provide an analysis of the sectoral impact of data science on SMEs business activities, with specific business use cases, and informing pertinent investors.
\item better understand the role that business associations, chambers of commerce, academia, national and local governments, international organisations, and other SMEs, could play to progress on these different dimensions, and support with knowledge sharing and open data availability.
\end{itemize}

Our analysis clearly suggests that most  SMEs collect  and store some sort  of business  data but require skills to analyse and  produce  useful  insights  for  data-driven  decision  making.  If  SMEs  are empowered  with  the  right  skills  and/or  supported  financially  for  this  purpose,  they  can  make  full utilization of data to help them, and hence driving the growth of entire economy.

% Numbered list
% Use the style of numbering in square brackets.
% If nothing is used, default style will be taken.
%\begin{enumerate}[a)]
%\item 
%\item 
%\item 
%\end{enumerate}  

% Unnumbered list
%\begin{itemize}
%\item 
%\item 
%\item 
%\end{itemize}  

% Description list
%\begin{description}
%\item[]
%\item[] 
%\item[] 
%\end{description}  

% Figure
%\begin{figure}[<options>]
%	\centering
%		\includegraphics[<options>]{}
%	  \caption{}\label{fig1}
%\end{figure}

%\begin{table}[<options>]
%\caption{}\label{tbl1}
%\begin{tabular*}{\tblwidth}{@{}LL@{}}
%\toprule
%  &  \\ % Table header row
%\midrule
% & \\
% & \\
% & \\
% & \\
%\bottomrule
%\end{tabular*}
%\end{table}

% Uncomment and use as the case may be
%\begin{theorem} 
%\end{theorem}

% Uncomment and use as the case may be
%\begin{lemma} 
%\end{lemma}

%% The Appendices part is started with the command \appendix;
%% appendix sections are then done as normal sections
%% \appendix

% To print the credit authorship contribution details
%\printcredits

%% Loading bibliography style file
%\bibliographystyle{model1-num-names}
%\bibliographystyle{cas-model2-names}
\Urlmuskip=0mu plus 1mu\relax
\bibliographystyle{elsarticle-num}

% Loading bibliography database
\bibliography{bibliography.bib}

% Biography
%\bio{}
% Here goes the biography details.
%\endbio

%\bio{pic1}
% Here goes the biography details.
%\endbio

\end{document}